\documentclass{optica-article}

\journal{opticajournal} 

\articletype{Research Article}

\usepackage{amsmath}           
\usepackage{booktabs}          
\usepackage{nicefrac}          
\usepackage{hyperref}          
\usepackage{url}               
\usepackage{orcidlink}         

\graphicspath{{figures/}}

\begin{document}

\title{Improving Richardson--Lucy Deconvolution with Diffusion Priors for Fluorescence Microscopy}

\author{Hao Chen\orcidlink{0009-0004-9430-2057}\authormark{1} and Scott S. Howard\orcidlink{0000-0003-3246-6799}\authormark{1,*}}

\address{\authormark{1}Department of Electrical Engineering, University of Notre Dame, Notre Dame, IN 46545, USA}

\email{\authormark{*}showard@nd.edu}

\begin{abstract*}
Richardson--Lucy (RL) deconvolution improves fluorescence microscopy images by recovering details lost to diffraction. It estimates the fluorescence signal most likely to have produced the measured photon counts under a Poisson imaging model. However, deconvolution remains ill-posed, especially at low photon counts, when weak biological structures become difficult to distinguish from shot noise and the prior strongly influences the reconstruction. RL recovers structure in early iterations but can amplify noise and become unstable, while regularizers such as total variation (TV) reduce these artifacts at the cost of oversmoothing. Instead, we learn a prior from real fluorescence microscopy images and use its gradient within an inverse-problem framework. The prior is estimated using an unconditional score-based diffusion model trained on a large microscopy dataset. At each step, the learned prior guides RL toward plausible specimen structures, while RL enforces Poisson consistency with the measured counts. Across diverse biological samples and cellular morphologies, the framework reduces RL noise amplification while better preserving weak structures at low photon counts.
\end{abstract*}

\section{Introduction}\label{sec:Intro}
Across biology and the life sciences, fluorescence microscopy makes labeled molecules such as proteins visible inside living cells, tissues, and whole organisms by forming an image from the light those labels emit~\cite{pawley2006handbook,Lichtman2005FluorescenceMicroscopy,lakowicz2006principles}. Modern modalities, including widefield, confocal, and two-photon microscopy, offer different trade-offs in field of view, optical sectioning, imaging depth, acquisition speed, and spatial resolution~\cite{denk1990two,zipfel2003nonlinear,Stemmer2008WidefieldFluorescence,Elliott2020ConfocalMicroscopy,conchello2005optical,helmchen2005deep}. Despite these differences, image formation in each case follows the same two physical steps. Light emitted by the specimen is first blurred by the point-spread function (PSF), the diffraction-limited intensity impulse response of the optical system. The blurred light is then registered as a finite number of photons, so the recorded image is a Poisson realization of the blurred intensity~\cite{Hell2003PSFEngineering,Zhang2007GaussianPSF,Boncelet2009ImageNoiseModels,Beenakker2003QuantumShotNoise}. Deconvolution techniques aim to recover the structure and density of the labeled specimen by inverting this image-formation model~\cite{Sibarita2005DeconvolutionMicroscopy,Bertero2021InverseProblemsImaging,Riad1986DeconvolutionProblem,Sarder2006Deconvolution3DFMM,Zunino2023ISMInverseProblem,Ribes2008LinearInverseImaging}.

Classical Richardson--Lucy (RL) deconvolution exactly inverts this model with expectation--maximization (EM) updates to maintain consistency under Poisson statistics~\cite{richardson1972bayesian,lucy1974iterative}. It has been widely applied in astronomy, optical microscopy, and 3D fluorescence microscopy~\cite{Dey2006RLTV,bertero2005simple,starck2002deconvolution,zhang2019super}. However, in the low-photon regime, RL iterations can amplify small perturbations into high-frequency artifacts~\cite{Ingaramo2014RichardsonLucy,Bi1994RLConvergence,Liu2025RLIllConvergence}. The PSF also discards information outright, attenuating the specimen's finest details at the objective in a way that no subsequent inversion can recover. Where the measurement no longer determines fine structure, the prior rather than the data determines the reconstruction. Classical practice supplies that prior as a preset structural constraint. Structure-based regularizers such as total-variation-regularized RL (TV-RL) impose a spatial smoothness assumption to damp oscillations and noise amplification~\cite{Dey2006RLTV,Zhang2026RLReview}. This stabilizes the iterations, but the imposed smoothness may suppress fine, weak subcellular structures and introduce prior-dependent artifacts. The strength of the regularization must also be retuned as the PSF and photon budget change.

Total variation regularization represents only one assumption about the structural constraints of fluorescence images. With deep learning, we can instead supply a learned distribution as the prior, estimated from real acquisitions using deep image priors~\cite{Ulyanov2020DeepImagePrior}, generative adversarial networks~\cite{Radford2016DCGAN,Zhu2017CycleGAN}, and score-based diffusion models~\cite{song2019generative,song2021scorebased,song2020improved,karras2022elucidating,Ho2020DDPM}, all of which have been used to constrain ill-posed inverse problems~\cite{chung2022diffusion,Zheng2025InverseBench}. For this study, we adopt a score-based diffusion model, the current state of the art in image generation. Trained to denoise images across dynamic noise levels, it estimates the gradient of the log-density of the learned image distribution, which is then combined with the measurement term at each step. Recent decoupled strategies keep that structural prior separate from the data-consistency optimization~\cite{zhang2025improving,Chen2025DAPSpp}, so the prior represents recurring cellular morphology, while the data-consistency step keeps each iteration consistent with the observed measurement. Yet most existing diffusion methods assume Gaussian noise, which can be conveniently paired with the learned prior, so their compatibility with Poisson photon statistics remains underexplored in biological imaging.

Here, we build a deconvolution method around a prior estimated from real fluorescence images, retaining the Poisson statistics of photon counting in the measurement term. The result is a decoupled diffusion-prior Richardson--Lucy (Diffusion-RL) framework for photon-limited deconvolution in microscopy. We estimate the prior by training an unconditional Elucidating Diffusion Models (EDM) architecture on diverse microscopy modalities and samples with varying cellular fields of view so that it captures recurring biological structures~\cite{karras2022elucidating}. Reconstruction then alternates between the two halves of the problem. Sampling the learned prior supplies a structural estimate, and RL refines that estimate against the PSF and measured photon counts, keeping each iterate consistent with Poisson detection. Structure therefore enters the reconstruction from the statistics of real specimens rather than from a generic assumption about images. This reduces the noise amplification of RL while retaining the weak subcellular structures, such as thin F-actin filaments and mitochondrial details, that a smoothness penalty suppresses. Both benefits hold across varying photon levels and imaging conditions with a single trained prior.

\section{Method}\label{sec:Method}
Inside the microscope, the PSF of the objective blurs the emitted light, and the limited photon budget makes its detection noisy; together, they form the measurement model. The morphology of labeled cellular structures, such as puncta and thin filaments, recurs across acquisitions and is a property of the specimen rather than of the microscope settings used to image it. Prior guidance and measurement consistency can therefore be separated, so a learned biological prior, pretrained without any PSF or photon-scale information, can be applied across microscope settings. Following a decoupled inference strategy~\cite{Chen2025DAPSpp}, the learned prior over morphology is evaluated without reference to the measured image, while the RL step enforces consistency among the PSF, Poisson statistics, and measured photon counts. Thus, the recovery of a clean image $\mathbf{x}$ from a blurred, photon-limited measurement $\mathbf{y}$ alternates between the prior and the RL update. We first specify the measurement model and the RL update with TV regularization and then extend this formulation to the decoupled diffusion solver.

\begin{figure}[htbp]
\centering
\includegraphics[width=0.85\linewidth]{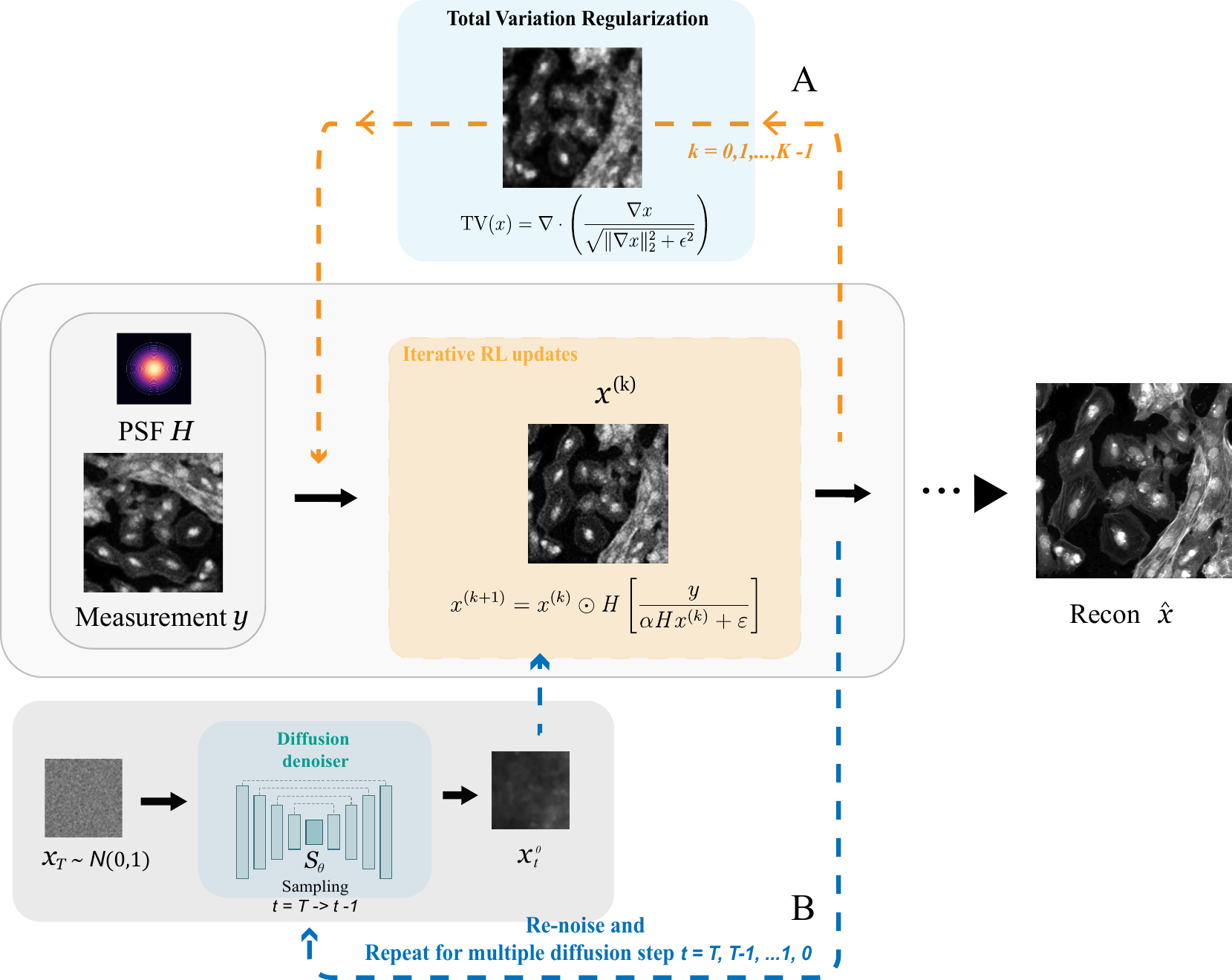}
\caption{\textbf{Overview of the Diffusion-RL framework.}
Classical Richardson--Lucy (RL) deconvolution iteratively restores an image using a known point-spread function (PSF). The two variants use
(A) total variation regularization and
(B) a diffusion prior that uses the diffusion model output as a warm initialization for RL, enabling repeated refinement through iterative deconvolution updates.}\label{fig:method}
\end{figure}

\subsection{Richardson--Lucy as a Poisson data-consistency step}\label{subsec:rl_likelihood}

For photon-limited fluorescence imaging, the measured photon counts are modeled as a Poisson process governed by PSF convolution and a photon scaling factor $\alpha$:
\begin{equation}
    \mathbf{y} \sim \mathrm{Poisson}\!\left(\alpha\,\mathbf{H}\mathbf{x}\right),
    \label{eq:poisson_observation_model}
\end{equation}
where $\mathbf{H}$ represents convolution with the microscope PSF, and the photon-count scale $\alpha$ sets the mean of the counts. Maximizing the corresponding Poisson log-likelihood under a physical nonnegativity constraint leads to the classical RL multiplicative update shown in Fig.~\ref{fig:method}(A):
\begin{equation}
    \mathbf{x}^{(k+1)} = \mathbf{x}^{(k)} \odot \left[ \mathbf{H}^{\ast} \frac{\mathbf{y}}{\alpha\,\mathbf{H}\mathbf{x}^{(k)} + \epsilon} \right],
    \label{eq:rl_update}
\end{equation}
where $\mathbf{H}^{\ast}$ is the adjoint convolution operator, $\odot$ denotes element-wise multiplication, and $\epsilon>0$ prevents division by zero. The update inverts that measurement model by maximum likelihood without any structural assumptions, so we leave the Poisson likelihood of RL unmodified and apply classical RL as a data-consistency operator inside the diffusion-prior sampling loop.

TV-RL optimizes the penalized negative log-likelihood objective $-\log p(\mathbf{y}\mid\mathbf{x}) + \lambda_{\mathrm{TV}}\|\nabla \mathbf{x}\|_{1}$, in which the total variation (TV) term penalizes the magnitude of the reconstruction's intensity gradient and thus favors fields that remain flat except at a few edges. The penalized objective leads to the following modified update:
\begin{equation}
    \mathbf{x}^{(k+1)} = \mathbf{x}^{(k)} \odot \frac{\mathbf{H}^{\ast} \left( \dfrac{\mathbf{y}}{\alpha\,\mathbf{H}\mathbf{x}^{(k)}+\epsilon} \right)}{1 - \dfrac{\lambda_{\mathrm{TV}}}{\alpha} \nabla\cdot \left( \frac{\nabla \mathbf{x}^{(k)}}{\sqrt{|\nabla \mathbf{x}^{(k)}|^2 + \epsilon_{\mathrm{TV}}^2}} \right)},
    \label{eq:tvrl_update}
\end{equation}
where $\lambda_{\mathrm{TV}}$ sets the weight of the smoothness term relative to the likelihood, and $\nabla\cdot(\cdot)$ is the discrete divergence. The divergence term in the denominator damps the high-frequency oscillations that photon noise drives in the unregularized update, but it does not distinguish weak subcellular structures from those oscillations. Because the penalty does not describe what fluorescence images look like, a punctum or thin filament whose intensity gradient is comparable to the noise is erased once $\lambda_{\mathrm{TV}}$ is large enough to control the noise.

\subsection{Score-based diffusion prior for fluorescence morphology}\label{subsec:decoupled_diffusion}
A learned prior instead uses the gradient of the log-density of fluorescence images $\nabla_{\mathbf{x}}\log p(\mathbf{x})$, called the score, to guide the reconstruction toward images consistent with real fluorescence data. We estimate it with a network trained under the Elucidating Diffusion Models (EDM) formulation~\cite{karras2022elucidating,song2019generative}, though any model returning a usable score could enter the reconstruction as a prior. In the forward perturbation process, independent and identically distributed (i.i.d.) isotropic Gaussian noise is added across a broad range of noise scales $\sigma_t$, producing noisy images $\mathbf{x}_t = \mathbf{x}_0 + \sigma_t \boldsymbol{\epsilon}$. A neural network $s_\theta(\mathbf{x}_t, \sigma_t)$ is trained to estimate the score function $\nabla_{\mathbf{x}_t}\log p_t(\mathbf{x}_t;\sigma_t)$, which describes how fluorescence images are distributed in space.

The measurement enters through Bayes' rule. Conditioning on the observed measurement $\mathbf{y}$ incorporates the PSF and the effects of limited photon counts, such that the posterior score decomposes into a prior term and a likelihood term:
\begin{equation}
    \nabla_{\mathbf{x}_t}\log p_t(\mathbf{x}_t \mid \mathbf{y};\sigma_t) = \nabla_{\mathbf{x}_t}\log p_t(\mathbf{x}_t;\sigma_t) + \nabla_{\mathbf{x}_t}\log p_t(\mathbf{y}\mid \mathbf{x}_t;\sigma_t).
    \label{eq:diffusion_posterior_score}
\end{equation}
The likelihood term is difficult to evaluate directly because doing so would require modeling the recorded photon counts conditioned on the intermediate noisy state $\mathbf{x}_t$. Under low photon budgets, shot noise is strongly signal-dependent and non-Gaussian, making likelihood gradients through $\mathbf{x}_t$ unreliable. We therefore decouple the prior and data-consistency steps, so the likelihood is evaluated only on the clean-image estimate~\cite{Chen2025DAPSpp}. At each sampling step, the prior network predicts a clean image estimate $\hat{\mathbf{x}}_0$. We then perform a separate, modular Bayesian data-consistency refinement directly on this clean projection:
\begin{equation}
\hat{\mathbf{x}}_0^{(j+1)} = \mathcal{B}_{\eta}\!\left( \hat{\mathbf{x}}_0^{(j)},\, \nabla_{\hat{\mathbf{x}}_0^{(j)}}\log p(\mathbf{y}\mid \hat{\mathbf{x}}_0^{(j)}) \right),
\label{eq:bayesian_dc_update}
\end{equation}
where $\mathcal{B}_{\eta}$ represents a likelihood operator specifically matched to the measurement statistics. Because the refinement is applied to a clean-image estimate rather than a noisy diffusion state, $\mathcal{B}_{\eta}$ can directly incorporate the prescribed measurement physics, independently of the generative prior learning process.

\subsection{Alternating diffusion-prior projection and RL data consistency}\label{subsec:diffrl}
We therefore instantiate $\mathcal{B}_{\eta}$ using RL as the data-consistency step in the decoupled framework, while the initial estimate is provided by the pretrained EDM network, as illustrated in Fig.~\ref{fig:method}(B). Each step along the deterministic reverse trajectory begins with the prior alone, which maps the perturbed state $\mathbf{x}_t$ to a clean structural estimate through Tweedie's denoising formula:
\begin{equation}
    \bar{\mathbf{x}}_{t}^{0}
    =
    \mathbf{x}_t + \sigma_t^2 s_\theta(\mathbf{x}_t,\sigma_t).
    \label{eq:tweedie_estimate}
\end{equation}
No information from the measured photons has entered this estimate, so the prior only initializes the inner RL refinement:
\begin{equation}
    \mathbf{x}_{t}^{(0)}
    =
    \bar{\mathbf{x}}_{t}^{0}.
    \label{eq:rl_initialization}
\end{equation}
For $k = 1,\ldots,K$, that estimate is refined against the recorded photon counts under the assumed PSF, using the update given in Eq.~\ref{eq:rl_update}.
After $K$ RL iterations, the data-consistent clean estimate is obtained as
\begin{equation}
    \hat{\mathbf{x}}_{t,\mathrm{RL}}^{0}
    =
    \mathbf{x}_{t}^{(K)}.
    \label{eq:rl_refined_clean}
\end{equation}

RL returns a clean image, but the reverse trajectory runs over noisy states, and the score is evaluated at a given noise level. The refined estimate is therefore re-noised back to the current level along the direction set by the polynomial scheduler:
\begin{equation}
    \hat{\mathbf{x}}_{t,\mathrm{RL}}
    =
    \hat{\mathbf{x}}_{t,\mathrm{RL}}^{0}
    +
    \sigma_t \hat{\boldsymbol{\epsilon}}_t.
    \label{eq:renoising}
\end{equation}

The state passed to the next noise level is therefore the re-noised $\hat{\mathbf{x}}_{i,\mathrm{RL}}$, so the later denoising steps produce estimates consistent with the given measurement.

When the noise level becomes sufficiently small, i.e., $\sigma_t < \bar{\sigma}$, the sampler applies an ordinary differential equation (ODE) solver to further recover fine-scale details consistent with the measurement:
\begin{equation}
    \mathbf{x}_{i+1}
    =
    \Phi
    \left(
        \hat{\mathbf{x}}_{i,\mathrm{RL}},\,
        \sigma_i,\,
        \sigma_{i+1}
    \right),
    \label{eq:ode_propagation}
\end{equation}
where $\Phi$ denotes the high-order EDM ODE update that propagates the re-noised RL-refined state from $\sigma_i$ to $\sigma_{i+1}$.

Therefore, the prior and the RL update play complementary roles: the prior proposes structures characteristic of fluorescence specimens, and RL corrects them against the recorded counts under the PSF. The correction step is fixed by the measurement because Poisson photon counting determines the likelihood, and that likelihood determines the RL update, while the proposal step need only return a clean estimate at the current noise level. Any model that learns the data distribution could take the place of the diffusion prior used here because the prior only has to propose structures, whereas RL enforces consistency with the physical measurement.

\subsection{Training data and network settings}\label{subsec:training_settings}

To build a robust prior capable of generalizing across various specimen types, we trained the foundational diffusion model unconditionally on a diverse collection of $125{,}464$ single-channel, $512{\times}512$ fluorescence micrograph patches with no paired measurements or reconstruction targets. The collection aggregates four public repositories spanning several illumination modalities and specimen classes: SR-Caco-2 (confocal molecular markers)~\cite{belharbi24-sr-caco-2}, the Fluorescence Microscopy Denoising (FMD) dataset (confocal, two-photon, and widefield illumination of cells and tissues)~\cite{zhang2019poisson}, the Neuronal Cells dataset (widefield cultures)~\cite{clissa2024fluorescent}, and RxRx1 (large-scale cell lines)~\cite{sypetkowski2023rxrx1}. All patches were rescaled to a common $[0,1]$ intensity range so that the network estimates a distribution over fluorescent structures rather than over the absolute intensity scale of the detector. For specimen-specific validation, we fine-tuned this foundational prior on a confocal BPAE dataset (mitochondrial, actin, and nuclear stains) from an out-of-focus microscopy correction study~\cite{zhang2022correction}, splitting the corpus into 240 images for fine-tuning and 60 for testing.

For both the foundational training and the BPAE fine-tuning, we estimated the score with the same denoising network: a U-Net with a base channel width of 64, a time-embedding dimension of 256, and channel multipliers of $(1,2,4,8,8)$. The noise scales follow the EDM parameterization: $\sigma_{\min}=0.002$, $\sigma_{\max}=80$, $\rho=7$, and a log-normal training noise schedule ($P_{\text{mean}}=-1.2, P_{\text{std}}=1.2$). The only parameter set from the images is the data scale, which is the intensity spread assumed by the loss weighting. Because fluorescence emission is confined to the labeled structures and most of the field is dark, we set it to $\sigma_{\text{data}}=0.4$ based on the variability across the training corpus. We ran both stages for 100 epochs with a batch size of 16 on four NVIDIA A100 GPUs, using AdamW (learning rate $5{\times}10^{-5}$ and weight decay 0.01) and an exponential moving average of the weights (decay 0.9999).

\subsection{Sampling and evaluation metrics}\label{subsec:sampling_settings}

During inference, the reverse trajectory is discretized into $N_t=100$ total sampling steps using a standard Poly-7 noise schedule. At each step, the inner RL consistency loop runs for $L=6$ iterations without early stopping. To moderate the strength of the data-consistency correction, we use a relaxation factor $\eta=0.1$, such that only $10\%$ of the RL correction is applied to the current estimate at each iteration.

Reconstructed images ($\hat{\mathbf{x}}$) are evaluated against their corresponding clean ground-truth (GT) images ($\mathbf{x}$) over a 150-image test set using the mean and standard deviation (SD) of the peak signal-to-noise ratio (PSNR) and structural similarity index (SSIM) to quantify field-averaged quality. Local recovery is then measured separately using the full width at half maximum (FWHM) of normalized one-dimensional intensity profiles $p(r)$ through isolated filaments and structural boundaries.

\section{Results}\label{sec:Res}

\subsection{Diffusion prior learns fluorescence morphology}\label{sec:edm_training}

To assess how well the pretrained diffusion model captures the data distribution of general fluorescence microscopy, we first sampled directly from the unconditional model before applying any deconvolution operator. Following deterministic diffusion sampling strategies~\cite{song2020ddim}, each sample was initialized from Gaussian noise $\mathbf{x}_{\sigma_0}\sim\mathcal{N}(0,\sigma_0^2\mathbf{I})$, and the sampler followed the probability-flow ODE with a 20-step Euler discretization:
\begin{equation}
    \hat{\mathbf{x}}_{0,i}
    =
    \mathbf{x}_{\sigma_i}
    +
    \sigma_i^2 s_\theta(\mathbf{x}_{\sigma_i},\sigma_i),
    \qquad
    \mathbf{x}_{\sigma_{i+1}}
    =
    \mathbf{x}_{\sigma_i}
    +
    (\sigma_{i+1}-\sigma_i)
    \frac{\mathbf{x}_{\sigma_i}-\hat{\mathbf{x}}_{0,i}}{\sigma_i},
    \label{eq:unconditional_euler_sampling}
\end{equation}

\begin{figure}[htbp]
\centering
\includegraphics[width=0.65\linewidth]{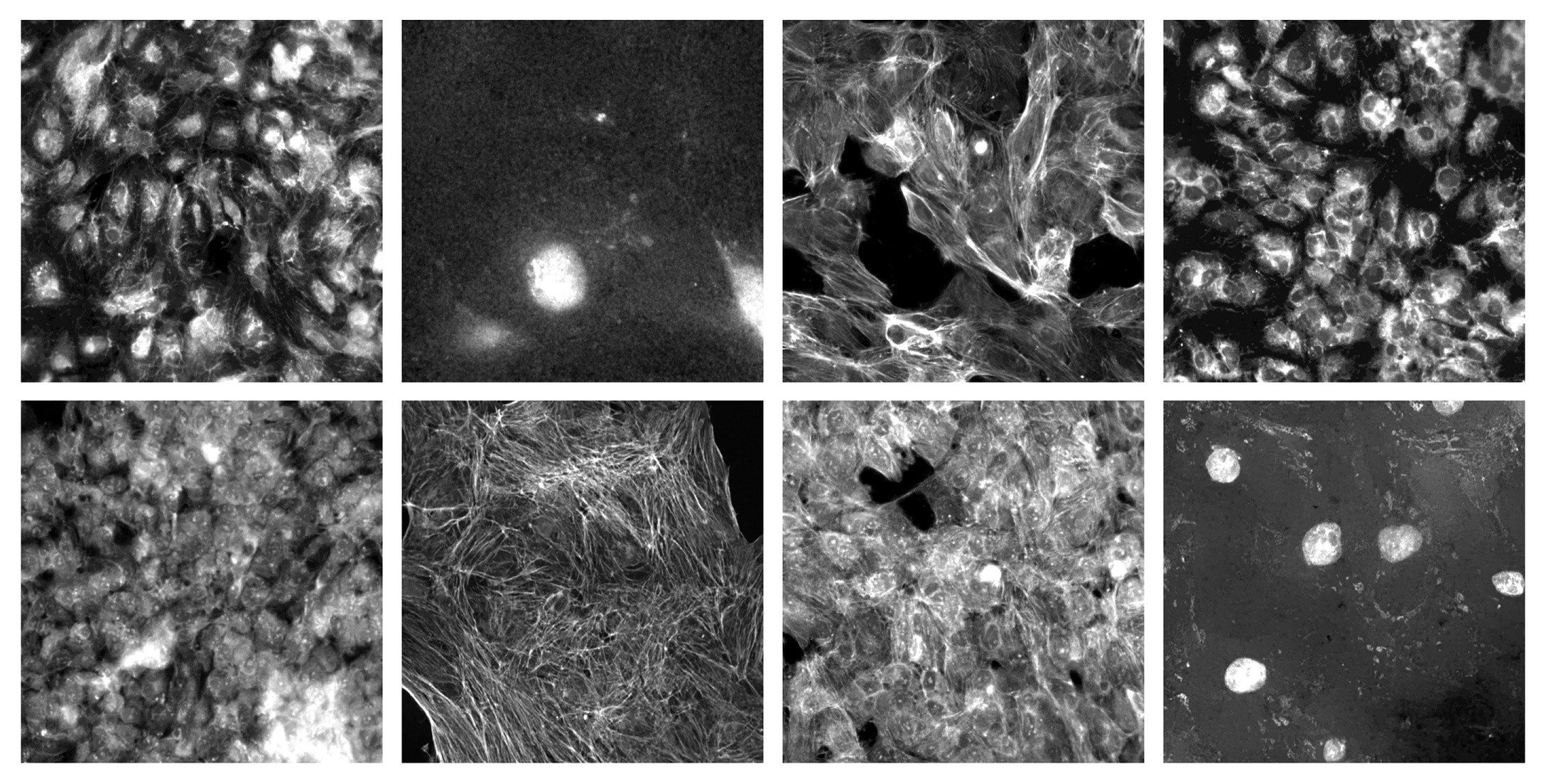}
\caption{\textbf{Unconditional diffusion sampling from the fluorescence prior.} Synthetic patches are generated from pure Gaussian noise using the pretrained unconditional diffusion model.}\label{fig:edm_uncond_samples}
\end{figure}
The final sample $\mathbf{x}_{\sigma_{20}}$ is therefore generated from the learned fluorescence prior alone and is controlled by the Gaussian-noise initialization, as shown in Fig.~\ref{fig:edm_uncond_samples}.

The unconditional samples contain compact puncta, filament-like networks, nuclear-like objects, and dense intracellular patterns, exhibiting a level of diversity similar to that observed in experimentally acquired fluorescence images. Once learned, this distribution serves as a pretrained prior for subsequent biological imaging applications~\cite{ma2024pretraining}. Here, the same prior is reused across different blur widths and photon levels rather than retrained for each imaging condition. However, unconditional samples alone cannot determine whether the learned distribution is a suitable prior for a specific measurement. This must be evaluated through measurement-conditioned reconstruction experiments.

\subsection{Diffusion-RL improves recovery of subcellular morphology under photon-limited imaging}\label{sec:diffusion_rl_results}
\begin{figure}[htbp]
\centering
\includegraphics[width=0.85\linewidth]{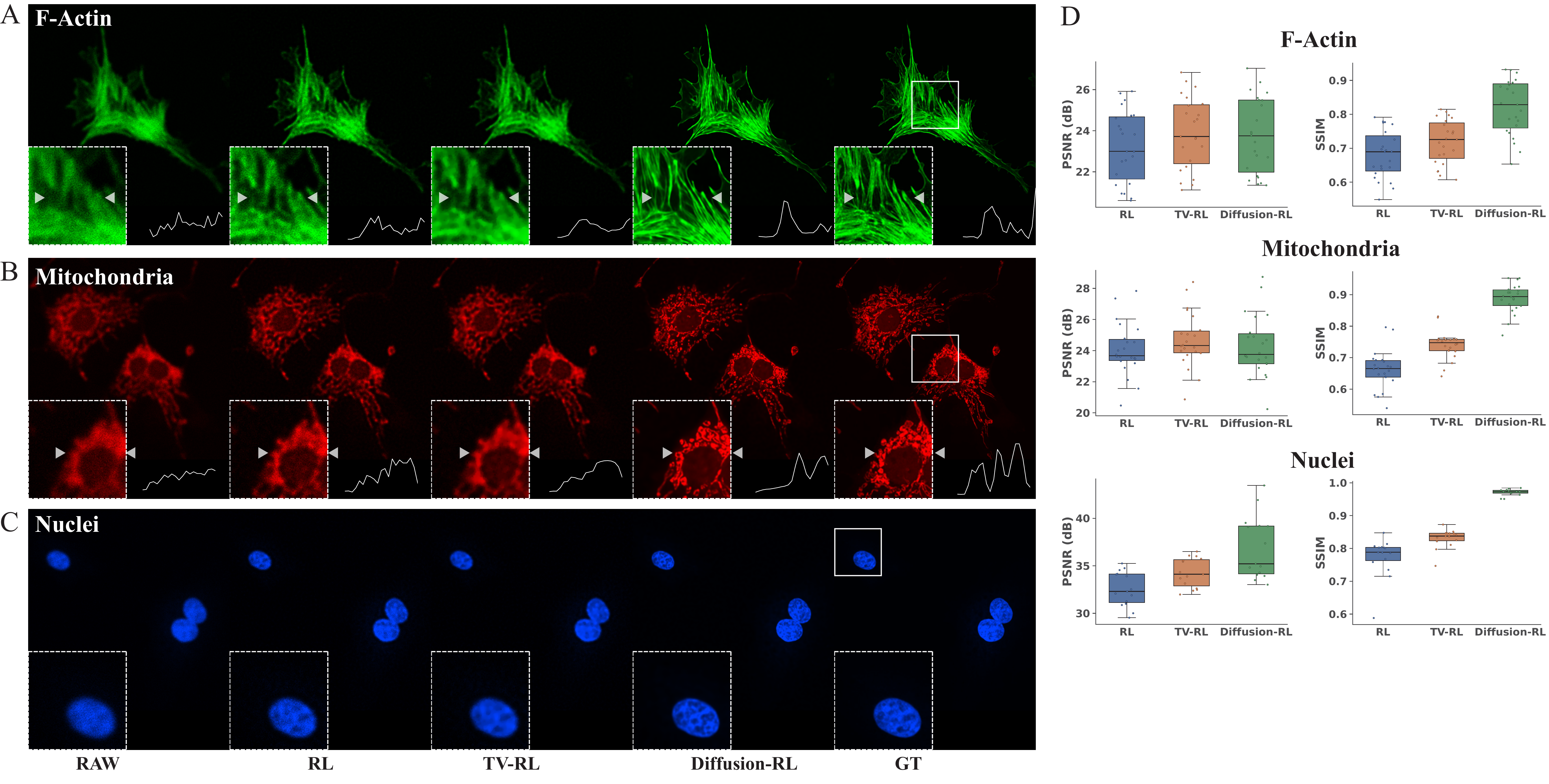}
\caption{\textbf{Comparisons on BPAE confocal fluorescence images.} Confocal fluorescence microscopy reconstructions of BPAE cells are shown for \textbf{(A)} F-actin (Alexa Fluor 488 phalloidin), \textbf{(B)} mitochondria (MitoTracker CMXRos, TxRed), and \textbf{(C)} nuclei (DAPI). For each target, the images compare the RAW measured input ($\alpha=100$), RL, TV-RL, Diffusion-RL, and the ground-truth (GT) reference. Insets for F-actin and mitochondria are evaluated using full width at half maximum (FWHM) measurements. \textbf{(D)} Statistical comparison of PSNR (dB) and SSIM across $N=20$ regions of interest.
}\label{fig:bpae_representative}
\end{figure}
The learned distribution is applied here to measurements that do not fully determine the underlying structure. Convolution with the PSF suppresses fine spatial details before detection, while photon-limited acquisition introduces Poisson noise, leaving weak subcellular features only weakly supported by the measurements. RL recovers the structure that can be explained by the observed photon counts and the assumed PSF, while the remaining ambiguity is resolved by the imposed prior. TV-RL introduces a smoothness prior that discourages undesirable variations in the reconstruction, whereas Diffusion-RL uses a learned distribution of fluorescence morphology to favor structures that are consistent with realistic fluorescence images while retaining the same Poisson data-consistency model enforced by RL. As the optics suppress finer structures and the photon budget becomes more limited, the reconstruction increasingly depends on the choice of prior.

To isolate the effect of the prior, measurements were simulated using Eq.~\eqref{eq:poisson_observation_model} with known blur kernels and photon budgets. The optical response was modeled using either a Gaussian PSF with $\sigma=3$\,px on a $31{\times}31$ support or a simulated optical PSF. Multiple photon-budget levels were considered to evaluate performance under different photon-counting conditions. For Diffusion-RL, we also used the foundational fluorescence prior fine-tuned on a small BPAE confocal dataset containing representative mitochondrial, F-actin, and nuclear structures~\cite{zhang2022correction}.

The reconstruction follows the decoupled update of Eq.~\ref{eq:bayesian_dc_update}, with RL serving as the Poisson data-consistency step. At each reverse-diffusion step, the diffusion prior produces a clean-image estimate $\hat{\mathbf{x}}_0$, which is refined by a small number of RL iterations before being passed to the next diffusion noise level. In this way, Diffusion-RL alternates between a diffusion step that promotes fluorescence-like structures consistent with the learned prior and an RL step that suppresses structures inconsistent with the measured photon counts.

Comparisons on BPAE confocal images are shown in Fig.~\ref{fig:bpae_representative} for F-actin labeled with Alexa Fluor 488 phalloidin, mitochondria labeled with MitoTracker CMXRos in the TxRed channel, and nuclei labeled with a blue-fluorescent DNA stain in the DAPI channel. The photon-limited input (RAW) retains only coarse cellular organization because the finest structures were already attenuated at the objective, and shot noise associated with the low photon counts corrupts what survives. RL sharpens the recovered structures but can also amplify photon noise. TV-RL suppresses this noise more effectively, although its smoothness prior can attenuate weak structures relative to the ground-truth (GT) reference. Diffusion-RL instead uses the learned fluorescence distribution to better preserve morphology across all three targets. F-actin filaments remain more distinct, mitochondrial structures appear more continuous, and nuclear details are better preserved within the nuclear boundaries.

Within each zoomed-in region, FWHM line profiles were measured across individual F-actin and mitochondrial structures. Diffusion-RL produces narrower profiles than those of the photon-limited input and RL while remaining closer to the ground-truth width. This agreement with the reference, rather than narrowing alone, indicates the recovery of structure rather than simple oversharpening. As the photon budget decreases, the learned prior plays an increasingly important role in the reconstruction, as explored in the following section.

\subsection{Reconstruction gives stronger gains at lower photon counts}\label{sec:photon_sweep_results}
Global reconstruction performance was quantified using \textbf{PSNR} and \textbf{SSIM}. The evaluation set comprised 150 fluorescence microscopy sub-images sampled from RxRx1~\cite{sypetkowski2023rxrx1} and SR-Caco-2~\cite{belharbi24-sr-caco-2}, with no overlap with the diffusion-prior training corpus.

The photon-count scale $\alpha$ was varied over four levels: low-count settings of 5 and 10 photons, an intermediate level of 100 photons, and a high-count setting of 255 photons. The quantitative comparison is reported in Table~\ref{tab:diffrl_compare}. Each measurement was formed as described above, with the ground-truth fluorescence image convolved with the assumed PSF and the convolved intensity then drawn as Poisson counts at the specified scale, so only the number of photons registered changes across the sweep.

\begin{table}[htbp]
\caption{Quantitative reconstruction comparison. Reconstruction methods are compared using PSNR (dB) and SSIM, reported as the mean$\,\pm\,$SD over 150 test images. Bold values indicate the best mean performance for each metric within each row.}
\label{tab:diffrl_compare}
\centering
\footnotesize
\setlength{\tabcolsep}{5pt}
\renewcommand{\arraystretch}{1.15}
\begin{tabular}{cc ccc}
\toprule
\multirow{2}{*}{PSF} & \multirow{2}{*}{$\alpha$} & RL & TV-RL & Diffusion-RL \\
 & & \scriptsize PSNR (dB)~/~SSIM & \scriptsize PSNR (dB)~/~SSIM & \scriptsize PSNR (dB)~/~SSIM \\
\midrule
\multirow{4}{*}{$31{\times}31,\sigma{=}3$}
& 5   & 14.86$\pm$2.14 / 0.247$\pm$0.172 & 15.72$\pm$2.10 / 0.259$\pm$0.171 &
\textbf{19.94$\pm$1.96} / \textbf{0.392$\pm$0.139} \\
& 10  & 17.24$\pm$2.21 / 0.306$\pm$0.167 & 18.79$\pm$2.13 / 0.335$\pm$0.161 &
\textbf{21.31$\pm$2.30} / \textbf{0.474$\pm$0.138} \\
& 100 & 24.18$\pm$2.64 / 0.582$\pm$0.140 & \textbf{26.17$\pm$3.02} / 0.672$\pm$0.131 &
25.93$\pm$3.45 / \textbf{0.714$\pm$0.147} \\
& 255 & 25.83$\pm$2.98 / 0.664$\pm$0.137 & \textbf{26.91$\pm$3.34} / 0.721$\pm$0.138 &
26.87$\pm$3.68 / \textbf{0.732$\pm$0.146} \\
\midrule
\multirow{4}{*}{$61{\times}61,\sigma{=}5$}
& 5   & 14.89$\pm$2.15 / 0.236$\pm$0.174 & 15.82$\pm$2.10 / 0.247$\pm$0.173 &
\textbf{20.17$\pm$1.97} / \textbf{0.402$\pm$0.138} \\
& 10  & 17.19$\pm$2.22 / 0.289$\pm$0.171 & 18.77$\pm$2.15 / 0.317$\pm$0.166 &
\textbf{21.63$\pm$2.40} / \textbf{0.480$\pm$0.135} \\
& 100 & 23.02$\pm$2.67 / 0.530$\pm$0.160 & 24.48$\pm$2.99 / 0.616$\pm$0.149 &
\textbf{24.71$\pm$3.38} / \textbf{0.665$\pm$0.163} \\
& 255 & 24.11$\pm$2.93 / 0.598$\pm$0.159 & 24.83$\pm$3.18 / 0.653$\pm$0.157 &
\textbf{25.05$\pm$3.48} / \textbf{0.671$\pm$0.162} \\
\midrule
\multirow{4}{*}{NA$=$1.05, $\lambda{=}650$nm}
& 5   & 14.56$\pm$2.13 / 0.241$\pm$0.161 & 15.27$\pm$2.10 / 0.252$\pm$0.159 &
\textbf{19.11$\pm$2.52} / \textbf{0.344$\pm$0.150} \\
& 10  & 16.91$\pm$2.19 / 0.306$\pm$0.157 & 18.26$\pm$2.11 / 0.334$\pm$0.151 &
\textbf{21.21$\pm$2.49} / \textbf{0.475$\pm$0.137} \\
& 100 & 24.69$\pm$2.52 / 0.620$\pm$0.119 & 27.07$\pm$2.88 / 0.710$\pm$0.107 &
\textbf{27.58$\pm$3.21} / \textbf{0.764$\pm$0.118} \\
& 255 & 26.98$\pm$2.90 / 0.719$\pm$0.111 & 28.50$\pm$3.35 / 0.777$\pm$0.113 &
\textbf{28.94$\pm$3.57} / \textbf{0.793$\pm$0.118} \\
\midrule
\multirow{4}{*}{\shortstack{Two-photon\\NA$=$0.8, $\lambda{=}800$nm}}
& 5   & 14.74$\pm$2.13 / 0.250$\pm$0.171 & 15.54$\pm$2.09 / 0.262$\pm$0.169 &
\textbf{20.04$\pm$2.10} / \textbf{0.388$\pm$0.137} \\
& 10  & 17.12$\pm$2.19 / 0.311$\pm$0.164 & 18.60$\pm$2.11 / 0.340$\pm$0.158 &
\textbf{21.89$\pm$2.36} / \textbf{0.502$\pm$0.132} \\
& 100 & 24.61$\pm$2.58 / 0.609$\pm$0.126 & 26.88$\pm$2.96 / 0.701$\pm$0.116 &
\textbf{27.53$\pm$3.33} / \textbf{0.757$\pm$0.127} \\
& 255 & 26.64$\pm$2.95 / 0.701$\pm$0.120 & 28.00$\pm$3.38 / 0.760$\pm$0.123 &
\textbf{28.48$\pm$3.61} / \textbf{0.777$\pm$0.127} \\
\bottomrule
\end{tabular}
\end{table}

\begin{figure}[htbp]
\centering
\includegraphics[width=0.85\linewidth]{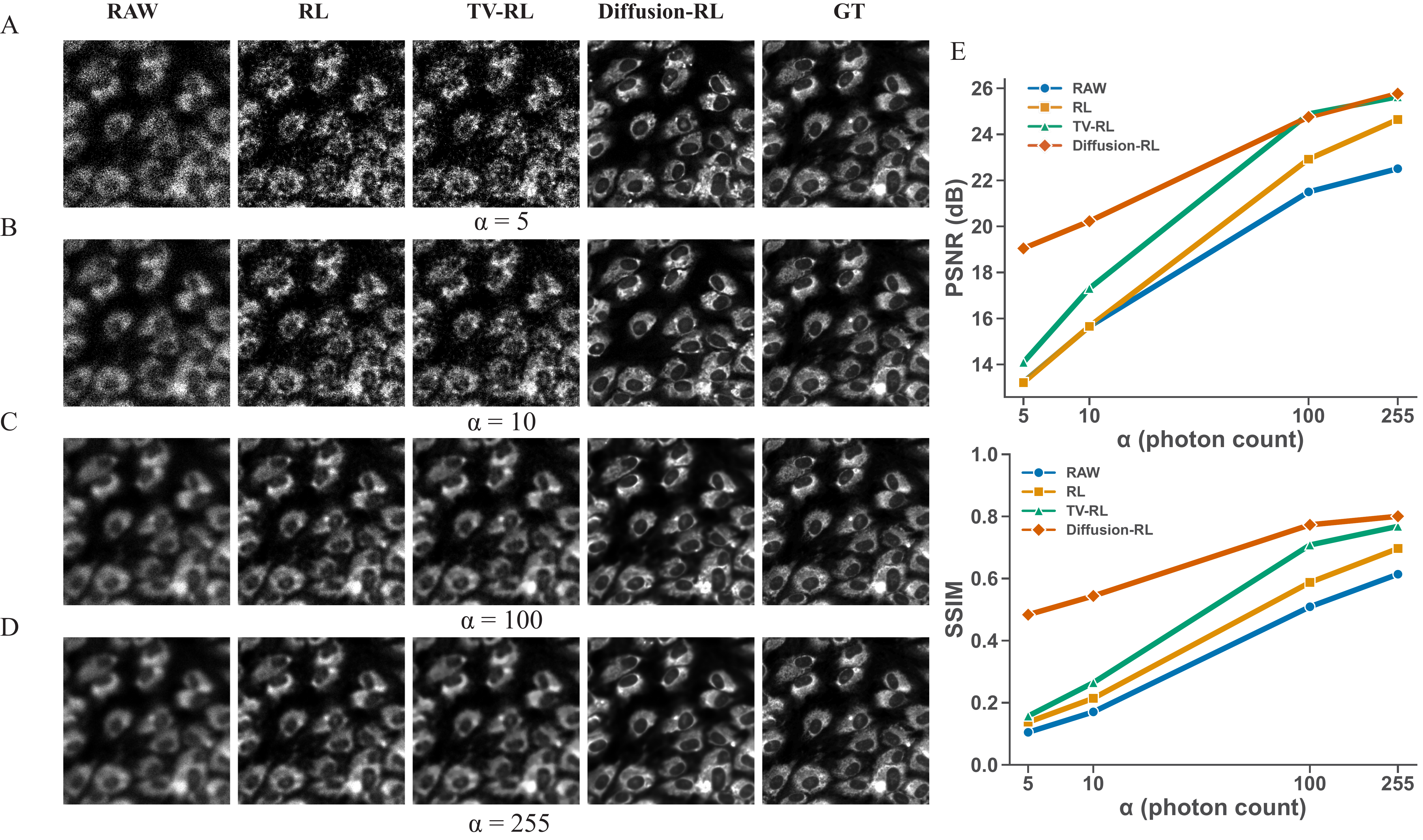}
\caption{\textbf{Reconstruction comparison at different photon-limited acquisition levels.} Reconstructions of widefield fluorescence images of HUVEC cells (ER/Nucleoli/RNA channel, 488~nm) under a 2D Gaussian PSF kernel are shown at \textbf{(A)} $\alpha=5$, \textbf{(B)} $\alpha=10$, \textbf{(C)} $\alpha=100$, and \textbf{(D)} $\alpha=255$. For each photon budget, the panels compare the RAW input, RL, TV-RL, Diffusion-RL, and the ground-truth (GT) reference. \textbf{(E)} PSNR (dB) and SSIM gains of each reconstruction method relative to the raw input across different photon budgets.}\label{fig:photon_sweep}
\end{figure}

Table~\ref{tab:diffrl_compare} shows that the performance margins are widest at $\alpha=5$ and $\alpha=10$, where Diffusion-RL leads on both metrics for every PSF setting considered. As the photon budget increases, these margins narrow. At $\alpha=100$ and $\alpha=255$, all three methods achieve reasonable global metrics. However, Diffusion-RL still recovers the sharpest and most recognizable fluorescence morphology, as shown in Fig.~\ref{fig:photon_sweep}. As $\alpha$ decreases, shot-noise fluctuations become increasingly large relative to the structure preserved after PSF blurring, making weak features difficult to distinguish from noise using the measurement alone. RL and TV-RL both struggle in this regime, but for different reasons. RL imposes no prior on specimen structure beyond nonnegativity, so its multiplicative update becomes increasingly driven by noise, and successive iterations can amplify these fluctuations. TV-RL instead favors piecewise-smooth reconstructions, which can suppress the weak intensity gradients that define faint biological features. In contrast, Diffusion-RL is the only method among the three that incorporates a learned model of plausible fluorescence structures, allowing it to better preserve weak features in the low-photon regime.

The contribution of the prior depends on how much uncertainty remains in the measured counts. When photons are plentiful, the data closely constrain the reconstruction, leaving little room for the prior to alter the result. When photons are scarce or the PSF is broad, many weak structural configurations remain consistent with the measurements, so the prior plays a larger role in selecting among them. This explains why the three methods diverge in the low-photon regime but converge as the photon budget increases. Diffusion-RL best preserves structural quality as the photon budget decreases because part of the reconstruction is guided by a learned prior rather than by the counts alone. The same score model is used across all photon levels, providing a consistent representation of fluorescence morphology. As a result, thin structures and local textures remain better resolved at low counts without the noise amplification seen in standard RL. This advantage is strongest in the lowest-photon regime, where the measurements constrain the reconstruction least strongly.

This low-count regime is directly relevant to fluorescence microscopy, where the photon budget is often limited by photobleaching, phototoxicity, acquisition speed, and live-cell viability. Under these measurement-limited conditions, faint fluorescence signals become comparable to shot-noise fluctuations and are further attenuated by diffraction. The diffusion prior provides structural guidance by supplying plausible fluorescence organization that the measured counts leave unresolved, while the RL step keeps the reconstruction consistent with the photons actually recorded. This combination helps preserve fine-scale structures that would otherwise be obscured by noise and optical degradation.

\subsection{Performance persists under different PSFs}\label{sec:psf_sweep_results}

\begin{figure}[htbp]
\centering
\includegraphics[width=0.85\linewidth]{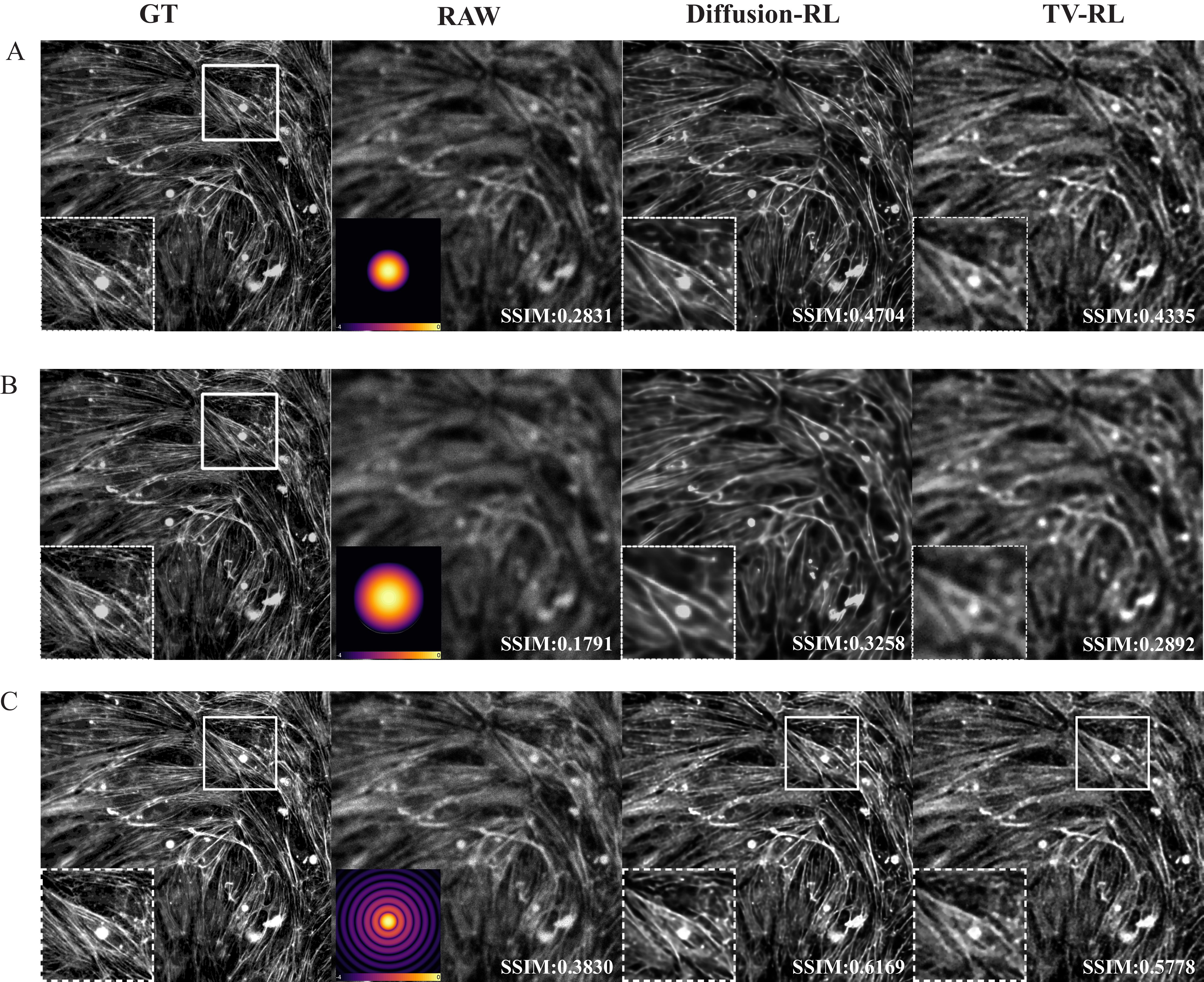}
\caption{\textbf{Reconstruction comparison under different PSF kernels.}
Reconstructions of widefield fluorescence images of HUVEC cells in the Actin/RNA channel at 488~nm are shown at a fixed photon budget of $\alpha=100$ for three PSF settings: \textbf{(A)} Gaussian PSF with $\sigma=3$~px and a $31{\times}31$ kernel; \textbf{(B)} Gaussian PSF with $\sigma=5$~px and a $61{\times}61$ kernel; and \textbf{(C)} simulated PSF with 650~nm excitation and NA~=~1.05. For each PSF kernel, the RAW input, TV-RL reconstruction, and Diffusion-RL reconstruction are compared. PSFs are displayed on a $\log_{10}$ scale.}
\label{fig:psf_sweep}
\end{figure}

A wider point-spread function attenuates more of the specimen’s high-spatial-frequency content, leaving less information about fine structures in the measurement. To evaluate this effect, we compare reconstructions at a fixed photon budget using two Gaussian PSFs, $\sigma=3$~px with a $31{\times}31$ kernel and $\sigma=5$~px with a $61{\times}61$ kernel, as shown in Fig.~\ref{fig:psf_sweep}. We also include a simulated confocal PSF to provide a more realistic optical model beyond the Gaussian approximation. As the PSF broadens, finer structural details are attenuated before detection, making the inverse problem increasingly ill-conditioned. Comparing RL, TV-RL, and Diffusion-RL across these settings therefore tests whether the learned fluorescence prior remains effective as progressively less structural information is preserved in the measurement.

As expected, the learned prior remains effective under the wider PSF, although reconstruction quality decreases for all methods. As reported in Table~\ref{tab:diffrl_compare}, the degradation is smallest for Diffusion-RL, which better preserves fluorescence morphology, local texture, and structural continuity than RL or TV-RL. Similar behavior under the simulated confocal PSF suggests that the benefit is not specific to Gaussian blur and can extend to other optical models when the corresponding PSF is provided. As the PSF broadens, fine structures are more strongly diffracted and become harder to distinguish, making deconvolution increasingly ill-conditioned. As a result, the optics impose a fundamental limit because recovery ultimately depends on how much structural information is preserved in the measurement, regardless of the prior provided.

\subsection{Reconstruction diversity reflects measurement-driven uncertainty}\label{sec:diversity_results}

\begin{figure}[htbp]
\centering
\includegraphics[width=0.85\linewidth]{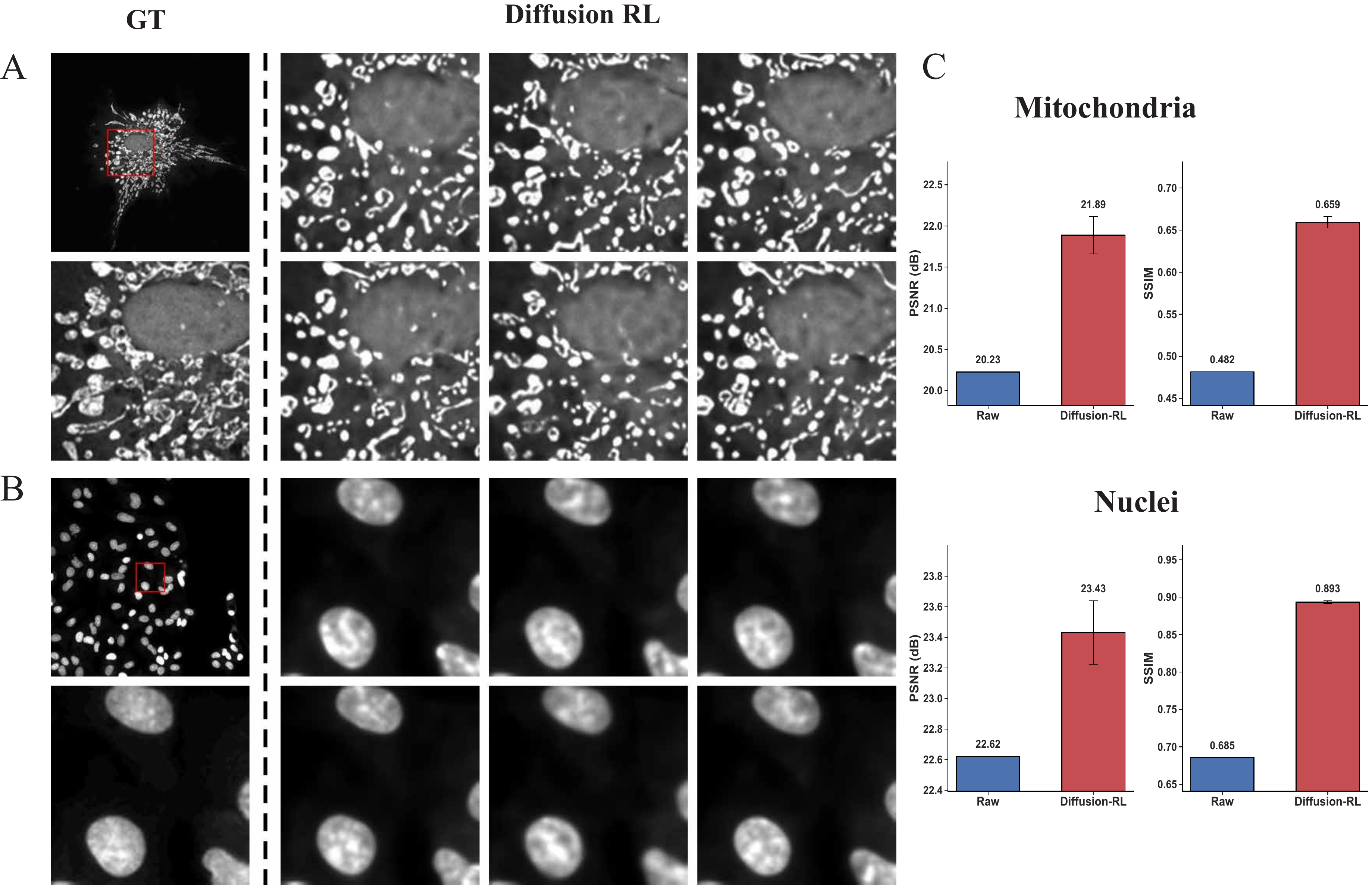}
\caption{\textbf{Diversity of Diffusion-RL reconstructions from stochastic sampling.} Multiple Diffusion-RL reconstructions are generated from different stochastic initializations for the same photon-limited measurement. The panels show \textbf{(A)} BPAE mitochondria and \textbf{(B)} HEPG2 nuclei (widefield fluorescence, 378~nm). \textbf{(C)} Sample-to-sample variation under the same measurement condition, with PSNR and SSIM quantified across $N=20$ samples.}\label{fig:diffusion_diversity}
\end{figure}

Beyond improving a single reconstruction, a generative prior can represent multiple plausible solutions to an ill-posed inverse problem. This is particularly useful in photon-limited microscopy, where PSF attenuation and low photon counts can leave weak or low-contrast structures insufficiently constrained by the measurement. In such cases, several arrangements of fine structure may remain consistent with the same recorded counts. Reconstruction diversity was therefore examined by drawing multiple reconstructions from different stochastic initializations for the same photon-limited observation, as shown in Fig.~\ref{fig:diffusion_diversity}. The PSF, photon budget, and score estimate were held fixed so that only the initialization varied, allowing the resulting spread to indicate where the measurement leaves the reconstruction uncertain.

Across these experiments, the contribution of the learned prior increases as the measurement becomes less informative. The photon sweep reduces information through lower photon counts, while the wider-PSF comparison reduces it through stronger optical attenuation. In both cases, Diffusion-RL combines the physical data consistency of RL with learned fluorescence morphology. The RL update keeps each reconstruction consistent with the measured photons and the known PSF, while the diffusion prior guides the reconstruction of features that are weakly constrained by shot noise or optical degradation. This leads to more stable and structurally faithful reconstructions than RL or TV-RL, particularly in low-photon-count and broader-PSF regimes, where the inverse problem is most ill-conditioned. These results show that Diffusion-RL is most beneficial when the measurement alone cannot reliably determine fine fluorescence structure.

\section{Discussion}\label{sec:Disc}
Diffusion-RL augments the Richardson--Lucy update with a prior learned from fluorescence images rather than one specified explicitly as a regularization penalty. The RL step enforces consistency with the recorded photon counts under the Poisson imaging model and the known PSF, while the reverse diffusion trajectory provides structural guidance from the learned image distribution. Although a diffusion model is used here, the framework only requires a generative model capable of producing a clean fluorescence estimate at the current noise level. As the photon budget decreases, the measurements constrain fine structures less strongly, increasing the influence of the learned prior on the reconstruction.

By decoupling prior learning from measurement physics, Diffusion-RL provides a modular alternative to end-to-end restoration networks. The learned fluorescence distribution contains neither the PSF nor the photon budget, since both are introduced only at inference through the RL likelihood step. In contrast, supervised methods such as CARE~\cite{weigert2018CARE}, 3D-RCAN~\cite{chen2021rcan3d}, Richardson--Lucy Net~\cite{li2022rln}, and deep light-sheet deconvolution networks~\cite{guo2020wbrl} learn a mapping from measurements to reconstructions using paired data matched to particular specimens, microscope configurations, and noise conditions. Changes in these conditions can therefore require new training pairs or retraining. Self-supervised approaches reduce the need for paired data but still embed degradation assumptions during training, such as pixel masking in Noise2Void~\cite{krull2019noise2void} or the re-blurring objective in ZS-DeconvNet~\cite{qiao2024zsdeconvnet}. In Diffusion-RL, the unconditional diffusion model instead acts as a PSF-independent morphology prior, while microscope-specific physics is supplied by RL at inference. Consequently, the same pretrained prior can be used across the PSF and photon regimes evaluated here without retraining.

The prior is a distribution rather than a fixed penalty, so Diffusion-RL can sample multiple admissible reconstructions from the same photon-limited measurement. Across different random seeds, the global cellular topology and major high-contrast structures remain stable, showing that the inner RL loop anchors each sample to the same measured photon counts and PSF. Variation is concentrated mainly in weak, low-contrast, or noisy regions where the measurement does not determine a unique structure. The resulting spread should therefore be interpreted as measurement-driven uncertainty rather than unconstrained hallucination. In contrast, a penalty-based method returns a single solution and does not reveal which structures are determined by the data and which are selected primarily by the prior.

Despite these advantages, several limitations remain. Diffusion-RL assumes a known, spatially invariant PSF, whereas real microscopy often exhibits depth- and position-dependent aberrations arising from wavefront distortions and refractive-index mismatch. Validation is also currently limited to two-dimensional images, while many fluorescence applications involve volumetric or time-lapse data. Computational cost is another limitation because repeated network evaluations along the reverse trajectory make Diffusion-RL substantially more expensive than classical regularized deconvolution. In addition, the sample-to-sample variation that reveals measurement uncertainty in weak structures must be interpreted carefully when reconstructions are used for quantitative analysis. Future work will therefore investigate blind or semi-blind PSF estimation, 3D diffusion priors, temporal consistency for live-cell imaging, and faster sampling or distillation strategies to reduce inference cost.

\section{Conclusion}\label{sec:Con}
We introduced Diffusion-RL, a decoupled Richardson--Lucy framework for photon-limited fluorescence deconvolution in which the prior is learned from fluorescence images rather than specified as a hand-designed penalty. Classical RL provides a physically grounded maximum-likelihood update based on the optical PSF and Poisson photon statistics, but without a strong structural prior, it can amplify noise and become unstable as photon counts decrease. Diffusion-RL addresses this limitation by using a diffusion model to represent a distribution of plausible fluorescence morphology, while the RL update enforces consistency with the measured photons and the known PSF throughout inference. The distinction between learned and conventional priors is small when photon counts are high and the blur is mild but becomes pronounced under low-photon-count and broader-PSF conditions, where the measurement no longer uniquely determines weak structures. In addition, stochastic sampling from the diffusion prior provides multiple physically consistent reconstructions, offering a practical view of measurement-driven uncertainty. Overall, Diffusion-RL shows how learned fluorescence morphology can strengthen classical Poisson deconvolution while preserving explicit dependence on the imaging physics and recorded data.


\bibliography{Diffusion_RL}

\end{document}